# Effect of residual strain on non-collinear antiferromagnetic structure in Weyl semimetal Mn$_3$Sn

J. J. Deng[1], J. Li[1], Y. Wang[1], X. Wu[1], X. T. Niu[1], L. Ma[1,a], D. W. Zhao[1], C. M. Zhen[1], D. L. Hou[1], E. K. Liu[2], W. H. Wang[2], and G. H. Wu[2]

[1]*Hebei Key Laboratory of Photophysics Research and Application, College of Physics, Hebei Normal University, Shijiazhuang 050024, China*
[2]*Beijing National Laboratory for Condensed Matter Physics, Institute of Physics, Chinese Academy of Sciences, Beijing 100190, China*

**ABSTRACT**

The non-collinear antiferromagnetic (AFM) structure makes Mn$_3$Sn exhibit exotic properties. At present, it has been found that both the hydrostatic pressure and the strain introduced by interstitial N atoms have a great influence on this magnetic structure. Here, the effect of the residual strain (RS) on it is investigated. AC and DC magnetic measurement results suggest that Mn$_3$Sn without RS has the non-collinear AFM structure only in the temperature range of 285 K to 400 K; while Mn$_3$Sn with RS has a non-coplanar AFM structure in the entire temperature range from 5 K to 400 K. Both anomalous Hall effect and topological Hall effect appears in Mn$_3$Sn with RS, supporting the anticipated non-coplanar AFM structure. Our findings point out a method to realize the chiral non-coplanar AFM structure through the engineering, thereby providing a path for the construction of topological antiferromagnets.

---

[a]Author to whom correspondence should be addressed: Electronic mail: majimei@126.com

# INTRODUCTION

Antiferromagnet has almost no stray field, and its spin dynamics is much faster than that of ferromagnet, therefore it has attracted great interest as a spintronic material for high-density and ultrafast memory devices.[1-3] Among them, the topological antiferromagnet $Mn_3Z$ (Z = Ge, Sn), the most striking of which,[4-7] exhibits large anomalous Hall effect,[8-10] anomalous Nernst effect,[11,12] large magneto-optical Kerr effect,[13] terahertz anomalous Hall effect,[14] planar Hall effect,[15,16] topological Hall effect[15,17-19] and other rich excellent properties.[20-23] These exotic properties originated from its unique non-collinear antiferromagnetic (AFM) structure.[24-26] In $Mn_3Z$, Mn atoms form a Kagome lattice, in which their spin present a 120° order with negative vector chirality, thus a non-collinear AFM structure appears.[27-30] This non-collinear AFM structure is derived from the competition of three interactions including Heisenberg exchange interaction, Dzyaloshinskii–Moriya (DM) interaction, and magnetocrystalline anisotropy.[27,31,32] Therefore, the non-collinear AFM structure of $Mn_3Z$ is a kind of dynamic equilibrium. The introduction of external fields such as temperature, stress or magnetic field can establish a new balance and a new chiral AFM structure. Temperature field makes non-collinear AFM structure[27-30] become non-coplanar AFM structure[33,34] or helical AFM structure[35,36] of $Mn_3Sn$. Hydrostatic pressure field causes a non-coplanar AFM structure or a collinear ferromagnetic structure of $Mn_3Ge$[37,38] and $Mn_3Sn$[39,40]. In addition, the strain introduced by interstitial N atoms has a great influence on the non-collinear AFM structure of $Mn_3SnN$.[41] Based on the above analysis, it is expected that the *residual strain* (RS) would also affect the non-collinear AFM structure of $Mn_3Z$.

In this paper, the Weyl semimetal $Mn_3Sn$ is selected as a research platform to explore the effect of the RS on the non-collinear AFM structure. AC and DC magnetic measurement results suggest that the RS causes the disappearance of the spin glass, the helical AFM structure and the non-collinear AFM structure, and the appearance of the frustrated AFM state and the non-coplanar AFM structure. The observations of anomalous Hall effect and topological Hall effect support the non-coplanar AFM structure in $Mn_3Sn$ with RS.

## EXPERIMENT DETAILS

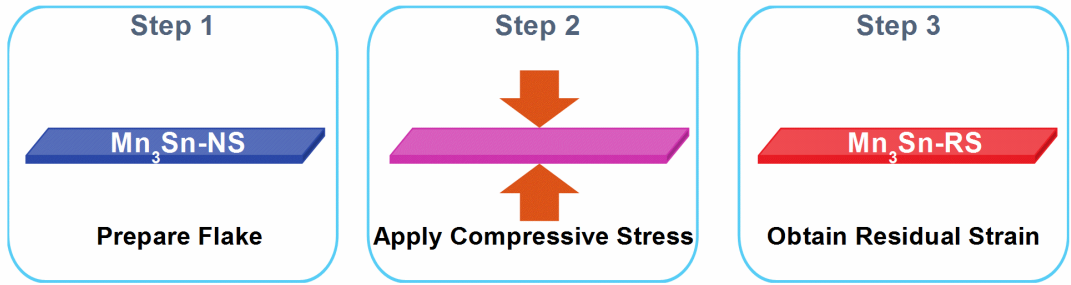

FIG. 1. Schematic diagram to obtain the residual strain in $Mn_3Sn$.

$Mn_3Sn$ polycrystalline sample was prepared by arc melting in an argon atmosphere, and the purity of elemental Mn and Sn were all above 99.99 %. In order to compensate for the loss of Mn during the arc melting process, approximately 3 wt. % of manganese was added. According to previous reports,[27,35] $Mn_3Sn$ is stable on rich Mn, so $Mn_{3.05}Sn_{0.95}$ is prepared. Based on the energy dispersive X-ray (EDX) spectroscopy result, the actual composition is $Mn_{3.04}Sn_{0.96}$ and the composition is uniform. For the convenience of description, the sample is named $Mn_3Sn$. Figure 1 shows a schematic diagram to obtain the sample of $Mn_3Sn$ with RS. Three steps are needed. First, $Mn_3Sn$ is cut into a flake with a size of 6×5×1.5 mm$^3$ by a sparker cutting machine. Next, the uniaxial compressive stress of 2 GPa is applied to the $Mn_3Sn$ flake for 10 minutes. Finally, the compressive stress is removed from the $Mn_3Sn$ flake sample, and $Mn_3Sn$ with RS is obtained. For ease of description, the $Mn_3Sn$ sample with RS is named $Mn_3Sn$-RS, and the $Mn_3Sn$ sample without strain is named $Mn_3Sn$-NS. The structure was determined by X-ray diffraction (XRD) with Cu $K_\alpha$ radiation. The DC magnetism of the sample is completed by the Magnetic Property Measurement System (MPMS, Quantum Design, Inc). The AC magnetic susceptibility and transport properties are measured by the Physical Property Measurement System (PPMS, Quantum Design, Inc).

## RESULTS AND DISCUSSION

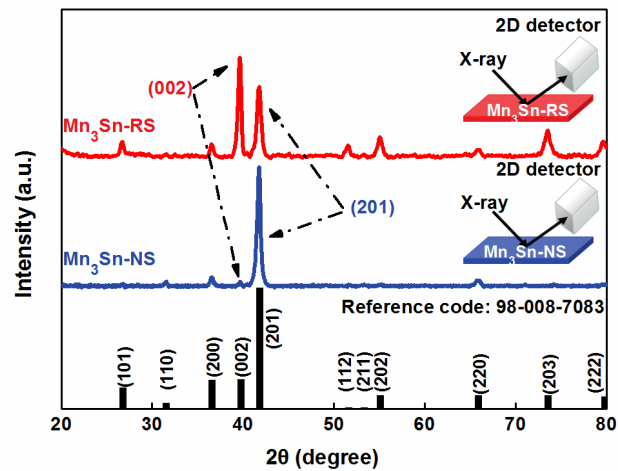

FIG. 2. XRD patterns of $Mn_3Sn$ without strain ($Mn_3Sn$-NS) and $Mn_3Sn$ with residual strain ($Mn_3Sn$-RS) at room temperature, respectively. The black bars are the standard peaks of the powder sample.

The insets are the schematic diagrams of the XRD measurement for $Mn_3Sn$-NS and $Mn_3Sn$-RS, respectively.

Figure 2 shows the XRD patterns of $Mn_3Sn$-NS and $Mn_3Sn$-RS at room temperature, respectively. By comparing the XRD of $Mn_3Sn$-RS with that of $Mn_3Sn$-NS, it can be seen that there are two influences of residual strain (RS) on the crystal structure. First, RS strongly enhances the intensity of (002) diffraction peak and weakens the intensity of (201) diffraction peak, as shown by the black dashed arrows in Fig. 2, indicating that RS reinforces the preferred orientation of the Kagome lattice. Second, RS makes the lattice parameters of $Mn_3Sn$ increase abnormally. The lattice parameters of $Mn_3Sn$-NS are $a$ = 5.6690 Å, $c$ = 4.5085 Å, V = 125.4787 Å$^3$, while those of $Mn_3Sn$-RS are $a$ = 5.7129 Å, $c$ = 4.5233 Å, V = 127.8494 Å$^3$. The RS values of lattice parameters are $\sigma_a$ = 0.77 %, $\sigma_c$ = 0.33 %, and $\sigma_V$ = 1.89 %, respectively. It is speculated that the positive RS values may be related to the change in the magnetic structure of $Mn_3Sn$.

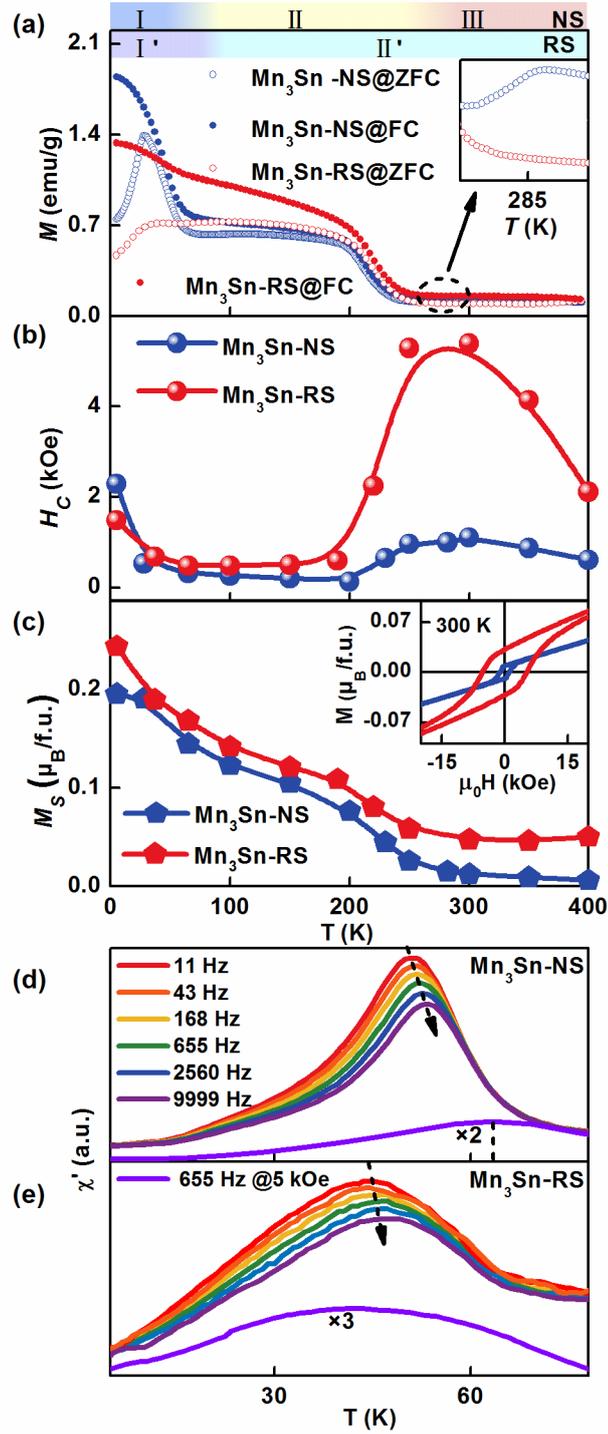

FIG. 3. Temperature dependence of the magnetization $M$ under 500 Oe field after zero field cooling (ZFC) and field cooling (FC) (a), the coercivity $H_C$ (b), the saturation magnetization $M_S$ (c), and the real part of the AC magnetic susceptibility $\chi'$ for Mn$_3$Sn without strain (Mn$_3$Sn-NS) (d) and Mn$_3$Sn

with residual strain (Mn₃Sn-RS) (e). Inset of (a) shows an enlargement of the black dashed circle. Inset of (c) shows the hysteresis loops for Mn₃Sn-NS and Mn₃Sn-RS at 300 K.

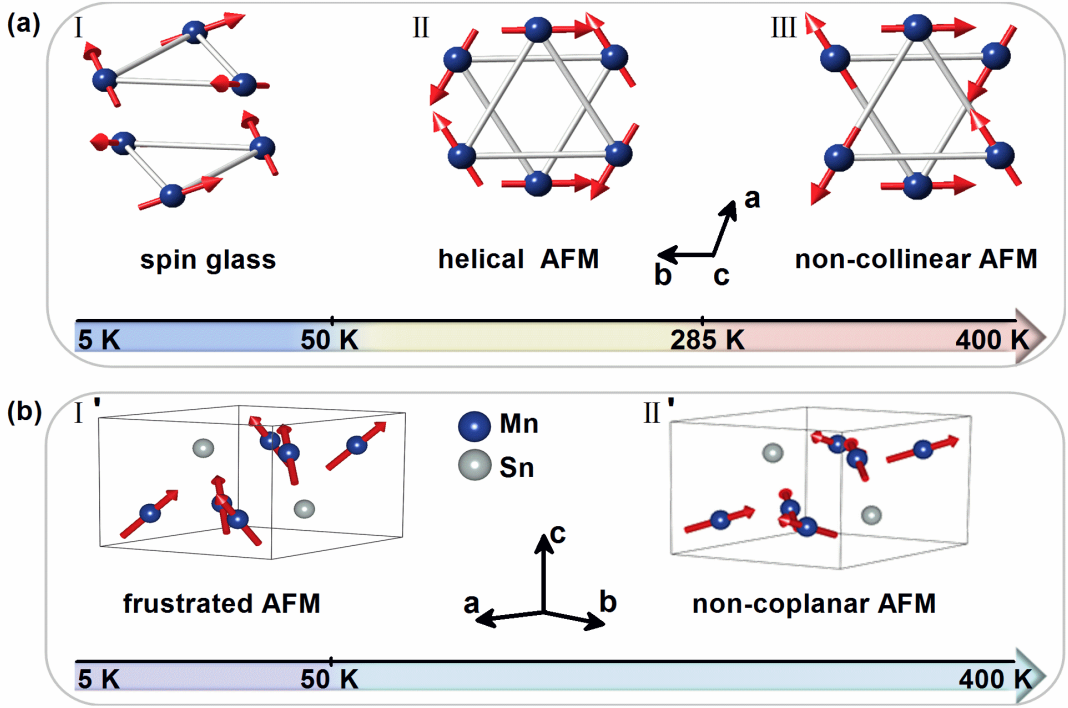

FIG. 4. Schematic diagrams of the magnetic structure temperature evolution for Mn$_3$Sn without strain (a) and Mn$_3$Sn with residual strain (b). AFM and red arrows represent antiferromagnetic and the direction of Mn moments, respectively.

Figure 3 shows the temperature dependence of the magnetization $M$ under 500 Oe field after zero field cooling (ZFC) and field cooling (FC), the coercivity $H_C$, the saturation magnetization $M_S$, and the real part of AC magnetic susceptibility $\chi'$ for Mn$_3$Sn-NS and Mn$_3$Sn-RS, respectively. For Mn$_3$Sn-NS, $M$, $H_C$ and $M_S$ all show three regions depending on temperature. For the convenience of expression, the three regions are called region I (from 5 K to 50 K), region II (from 50 K to 285 K) and region III (from 285 K to 400 K). In region III, it is seen that both $M$ and $M_S$ exhibit near-zero values [Figs. 3(a) and 3(c)], and $H_C$ is ~ 1000 Oe [Fig. 3(b)], which is consistent with the characteristic of the non-collinear AFM structure [Fig. 4(a III)]. In region II, a sudden change at 285 K is observed from $M(T)$ (the blue curve in the inset of Fig. 3(a)), and the $H_C$ also decays sharply to approximately zero [Fig. 3(b)]. These phenomena satisfy a helical AFM structure [Fig. 4(a II)].[36,42] In region I, the DC-ZFC curve has a peak [Fig. 3(a)], and the AC-$\chi'$ curve also has a peak whose position shows a strong frequency dependence [Fig. 3(d)], indicating the spin glass state [Fig. 4(a I)].[33,34] Therefore, the temperature evolution of the magnetic structure of $Mn_3Sn$-NS can be expected, as shown in Fig. 4(a), which shows its schematic diagram. Our speculation is consistent with the Refs 36,43,44.

Compared with $Mn_3Sn$-NS, the RS causes three changes in magnetic properties. First, both $H_C$ and $M_S$ of $Mn_3Sn$-RS are much higher than those of $Mn_3Sn$-NS in temperature region from 50 K to 400 K [Figs. 3(b) and 3(c)]. The inset of Fig. 3(c) shows the typical hysteresis loops for $Mn_3Sn$-NS and $Mn_3Sn$-RS. The $H_C$ and $M_S$ values of $Mn_3Sn$-RS are 5400 Oe and 0.05 $\mu_B$/f.u., respectively, which are 4 times of those of $Mn_3Sn$-NS. Therefore, it is speculated that RS makes the Mn moments in the Kagome lattice tilt toward the $c$ axis as shown in Fig. 4(b II'). Second, different from the case of $Mn_3Sn$-NS, the transition from the non-collinear AFM structure to the helical AFM structure is not observed in the inset of Fig. 3(a), which confirmed that there is no helical AFM structure in $Mn_3Sn$-RS. That is, with the help of RS, from 50 K to 400 K the non-coplanar AFM structure replaces the non-collinear AFM structure or the helical AFM structure. This phenomenon is very similar to the case of $Mn_3Ge$ hydrostatic pressure.[37,38] Third, below 50 K the frequency shift of the peak for $Mn_3Sn$-RS in Fig. 3(e) becomes weak, and the peak position shows robust characteristics to DC bias field up to 5 kOe. Thus, the RS changes the frustrated characteristics of $Mn_3Sn$ at low temperature. Figure 4(b) shows the temperature evolution from a non-coplanar AFM structure [Fig. 4(b II')] to a frustration AFM state [Fig. 4(b I')] in $Mn_3Sn$-RS.

In addition, for both $Mn_3Sn$-NS and $Mn_3Sn$-RS, Figs. 3(a) and 3(c) show a jump at 240 K, which is exactly the Curie temperature of $Mn_2Sn$. However, $Mn_2Sn$ phase is not observed in XRD patterns. It is speculated that there may be a very small amount of $Mn_2Sn$ remaining in the $Mn_3Sn$ sample.[35,45] The above-mentioned magnetic structures of $Mn_3Sn$-NS and $Mn_3Sn$-RS are only deduced from macroscopic magnetic measurements. Its correctness still needs to be confirmed by other experiments such as neutron diffraction, transport properties, etc.

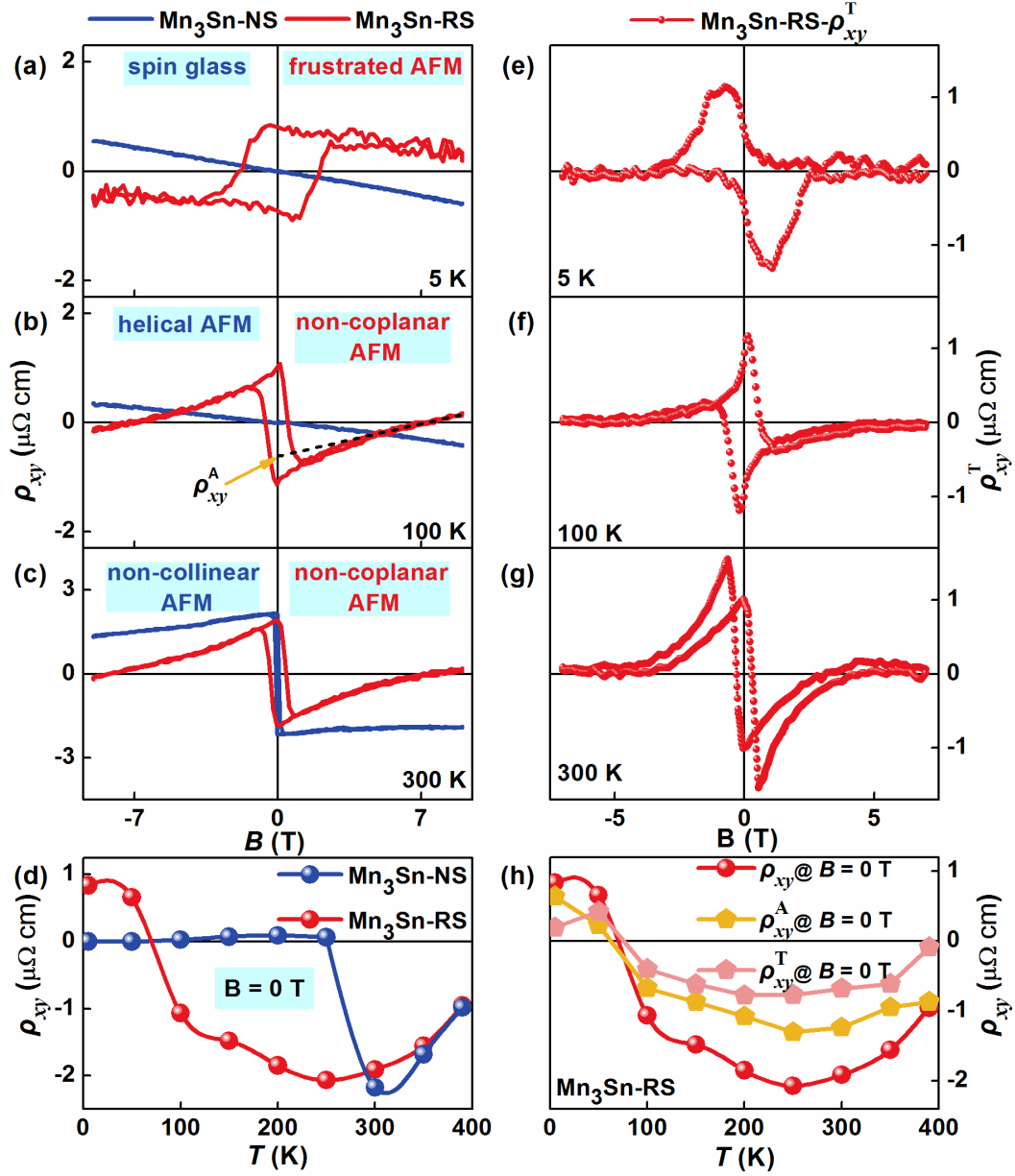

FIG. 5. Hall resistivity $\rho_{xy}$ as a function of magnetic field at 5 K (a), 100 K (b), and 300 K (c) for Mn$_3$Sn without strain (blue curves) and Mn$_3$Sn with residual strain (red curves). The black dashed line is the high-field linear extrapolation of $\rho_{xy}(B)$ to obtain the anomalous Hall resistivity $\rho_{xy}^A$. (d) Temperature dependence of $\rho_{xy}$ under B = 0 T for Mn$_3$Sn without strain and Mn$_3$Sn with residual strain. Topological Hall resistivity $\rho_{xy}^T$ as a function of magnetic field at 5 K (e), 100 K

(f), and 300 K (g) for Mn$_3$Sn with residual strain. (h) Temperature dependence of $\rho_{xy}$, $\rho_{xy}^{A}$ and $\rho_{xy}^{T}$ at B = 0 T for Mn$_3$Sn with residual strain.

Figures 5(a)-5(c) show the Hall resistivity $\rho_{xy}$ as a function of magnetic field at typical temperatures for Mn$_3$Sn-NS and Mn$_3$Sn-RS, whose magnetic structures at the corresponding temperature deduced in Fig. 4 are marked with blue and red fonts, respectively. It is seen that Mn$_3$Sn-NS exhibits a linearly reversible ordinary Hall effect (OHE) at 5 K (region I, spin glass) and 100 K (region II, helical AFM), and an anomalous Hall effect (AHE) at 300 K (region III, non-collinear AFM). Neither spin glass nor helical AFM structure breaks the time reversal symmetry,[36] thus only the OHE is observed [Figs. 5(a) and 5(b)];[43,44] while the non-collinear AFM structure breaks the time reversal symmetry and generates a fictitious magnetic field in momentum space,[24,26] resulting in AHE [Fig. 5(c)].[8] Different from the case of Mn$_3$Sn-NS, Mn$_3$Sn-RS shows AHE at all three typical temperatures. Moreover, the $\rho_{xy}(B)$ curve at 5 K (region I', frustration AFM) is in the first and third quadrants, while at 100 K and 300 K (region II', non-coplanar AFM) it is in the second and fourth quadrants. Based on the second part of formula $\frac{d\langle \vec{r} \rangle}{dt} = \frac{\partial E}{\hbar \partial \vec{k}} + \frac{e}{\hbar} E \times b_n$,[46] it is speculated that it may be caused by the change in the Berry curvature which originates from the magnetic structure.[24,26] That is, below 50 K, the Mn moments in Mn$_3$Sn would tilt toward $c$ axis more seriously than the non-coplanar AFM structure.[33,34] And the frustrated AFM state appears. In Fe$_3$Sn$_2$ and Co$_3$Sn$_2$S$_2$, the magnetic moments completely parallel to the $c$ axis, their $\rho_{xy}(B)$ curves are in the first and third quadrants.[47,48] Figure 5(d) shows the temperature dependence of Hall resistivity $\rho_{xy}$ under B = 0 T for Mn$_3$Sn-NS and Mn$_3$Sn-RS. It can be seen that Mn$_3$Sn-RS shows a finite spontaneous $\rho_{xy}$ in the whole temperature range of 5 K to 400 K, while Mn$_3$Sn-NS only exhibits the spontaneous $\rho_{xy}$ above 300 K. Besides, the spontaneous $\rho_{xy}$ values of Mn$_3$Sn-NS and Mn$_3$Sn-RS are almost the same near room temperature. Therefore, the RS expands the temperature window of the spontaneous $\rho_{xy}$ while keeping the spontaneous $\rho_{xy}$ value unchanged at high temperature.

The black dashed line in Fig. 5(b) is the high-field linear extrapolation of $\rho_{xy}(B)$, whose intersection with $\rho_{xy}$ axis is the anomalous Hall resistivity $\rho_{xy}^{A}$. Interestingly, it is found that the spontaneous $\rho_{xy}$ value is larger than that of $\rho_{xy}^{A}$, which indicates that there should be a contribution from the topological Hall resistivity $\rho_{xy}^{T}$ originating from the non-coplanar AFM structure.[49] The $\rho_{xy}^{T}$ in Mn$_3$Sn-RS is obtained by using the method in Refs. 17. Figures 5(e)-5(g) show $\rho_{xy}^{T}(B)$ curves at typical temperatures for Mn$_3$Sn-RS. It can be seen that the topological Hall effect (THE) appears at all typical temperatures, suggesting the nonzero scalar spin chirality from the non-coplanar AFM structure in Mn$_3$Sn-RS.[50] Refs. 15,17,18,19,44 also reported the THE in Mn$_3$Sn. There are two views on its mechanism, including the chiral domain walls[15,17,44] and the non-coplanar AFM structures[18,19]. Considering that the $M_S$ value of Mn$_3$Sn-RS is much larger than that of Mn$_3$Sn-NS, the THE in Mn$_3$Sn-RS should be caused by the non-coplanar AFM structure rather than the chiral domain wall. Fig. 5(h) shows the temperature dependence of $\rho_{xy}$, $\rho_{xy}^{A}$ and $\rho_{xy}^{T}$ at B = 0 T for Mn$_3$Sn-RS. It can be seen that both $\rho_{xy}^{A}$ and $\rho_{xy}^{T}$ exist in the entire temperature range from 5 K to 400 K, and their changing trends are almost the same. Ref. 24,26 point out that the non-coplanar AFM structure in Mn₃Z (Z = Ge, Sn) can give rise to not only AHE but also THE. In our case, the residual strain brings the non-coplanar AFM structure to Mn₃Sn, thus makes it exhibit both AHE and THE.

The above transport results for Mn₃Sn-NS and Mn₃Sn-RS support the expected magnetic structure. Residual strain can be an effective method for Mn₃Sn to construct a non-coplanar AFM structure.

## CONCLUSIONS

In conclusion, the effect of the residual strain (RS) on the magnetic structure of Mn₃Sn has been studied by the AC and DC magnetic properties and the Hall transport measurements. The experimental results consistently reflect the changes in the magnetic structure of Mn₃Sn caused by the RS in different temperature regions. In the high temperature region from 285 K to 400 K, with the help of RS the non-coplanar antiferromagnetic (AFM) structure replaces the non-collinear AFM structure, thus both anomalous Hall effect (AHE) and topological Hall effect (THE) are simultaneously observed in Mn₃Sn. In the middle temperature region from 50 K to 285 K, RS suppresses the appearance of the helical AFM state which cannot break the time-reversal symmetry and retains the non-coplanar AFM structure. In the low temperature region below 50 K, RS induces the frustration AFM state which tilts toward *c* axis more severely than the non-coplanar AFM structure, consequently the $\rho_{xy}(B)$ curve is in the first and third quadrants. The present study provides a method to construct a non-coplanar AFM structure in non-collinear AFM materials through strain engineering.

## ACKNOWLEDGEMENTS

This work was supported by the National Natural Science Foundation of China (Grant No. 11504247, 51901067 and 51971087), the Natural Science Foundation of Hebei Province (Grant No. A2018205144, A2017210070, E2016205268 and E2019205234), the Key Project of Natural Science of Hebei Higher Education (Grant No. ZD2017045), the Science and Technology Research Project of Hebei Higher Education (Grant No. QN2019154), and the Science Foundation of Hebei Normal University (Grant No. L2019B11).

## DATA AVAILABILITY

The data that support the findings of this study are available from the corresponding author upon reasonable request.